\documentclass[10pt,aps,prx,amsfonts,amsmath,amssymb,raggedbottom,longbibliography,reprint,superscriptaddress,citeautoscript,nofootinbib]{revtex4-2}
\usepackage[usenames,dvipsnames]{color}
\usepackage{graphicx,microtype}
\usepackage[bookmarks=false,colorlinks]{hyperref}
\usepackage{xspace}
\hypersetup{linkcolor=Plum,citecolor=Plum,filecolor=Plum,urlcolor=PineGreen}
\usepackage[normalem]{ulem}

\newcommand{\CNO}{Cd$_2$Nb$_2$O$_7$\xspace}
\newcommand{\GMIV}{$\Gamma_4^-$\xspace}
\newcommand{\GMV}{$\Gamma_5^-$\xspace}

\begin{document}

\title{Local structure and its implications for the relaxor ferroelectric Cd$_2$Nb$_2$O$_7$}

\author{Daniel Hickox-Young}
\affiliation{Department of Materials Science and Engineering, Northwestern University, Evanston, IL 60208, USA}

\author{Geneva Laurita}
\affiliation{Department of Chemistry and Biochemistry, Bates College, Lewiston, ME 04240, USA}

\author{Quintin N. Meier}
\affiliation{Université Grenoble Alpes, CEA, LITEN, 17 rue des Martyrs, 38054 Grenoble, France}

\author{Daniel Olds}
\affiliation{National Synchrotron Light Source II, Brookhaven National Laboratory, Upton, NY 11973, USA}

\author{Nicola A. Spaldin}
\affiliation{Materials Theory, ETH Zurich, Wolfgang-Pauli-Strasse 27, 8093 Z\"urich, Switzerland}

\author{Michael R. Norman}
\email{norman@anl.gov}\affiliation{Materials Science Division, Argonne National Laboratory, Lemont, IL  60439, USA}

\author{James M. Rondinelli}
\email{jrondinelli@northwestern.edu}\affiliation{Department of Materials Science and Engineering, Northwestern University, Evanston, IL 60208, USA}
\affiliation{Northwestern-Argonne Institute of Science and Engineering (NAISE), Northwestern University, Evanston, IL 60208, USA}

\date{\today}

\begin{abstract}
The relaxor ferroelectric transition in \CNO is thought to be described by the unusual condensation of two $\Gamma$-centered phonon modes, \GMIV and \GMV. However, their respective roles have proven to be ambiguous, with disagreement between \textit{ab initio} studies, which favor \GMIV as the primary mode, and global crystal refinements, which point to \GMV instead. 
Here, we resolve this issue by demonstrating from x-ray pair distribution function measurements that locally, \GMIV dominates, but globally, \GMV dominates.  This behavior is consistent with the near degeneracy of the energy surfaces associated with these two distortion modes found in our own \textit{ab initio} simulations. 
Our first-principles calculations also show that these energy surfaces are almost isotropic, providing an explanation for the numerous structural transitions found in \CNO, as well as its relaxor behavior. 
Our results point to several candidate descriptions of the local structure, some of which demonstrate two-in/two-out behavior for Nb displacements within a given Nb tetrahedron.  Although this suggests the possibility of a charge analog of spin ice in \CNO, our results are more consistent with a Heisenberg-like description for dipolar fluctuations rather than an Ising one.  We hope this encourages future experimental investigations of the Nb and Cd dipolar fluctuations, along with their associated mode dynamics.
\end{abstract}

\maketitle

\section{Introduction}

Relaxor ferroelectrics are a fascinating class of materials \cite{Cowley2011} whose dielectric properties depend strongly on frequency, with dynamics described by a Vogel-Fulcher relation, a relation also found for the magnetic response in spin-ice materials \cite{Kassner2015,Eyvazov2018,samarakoon2021structural}.
They are typically chemically complex solid solutions of perovskites, with the relaxor behavior thought to be driven by the dynamics of local polar clusters.  Yet the disorder inherent in these  materials inhibits unambiguous interpretation of the data, leading to differing ideas about the origin of relaxor behavior.  Therefore, there has been a search for stoichiometric materials that exhibit relaxor behavior that would aid in identifying intrinsic behavior and help settle this debate.

\begin{figure*}
\centering
\includegraphics[width=1.72\columnwidth]{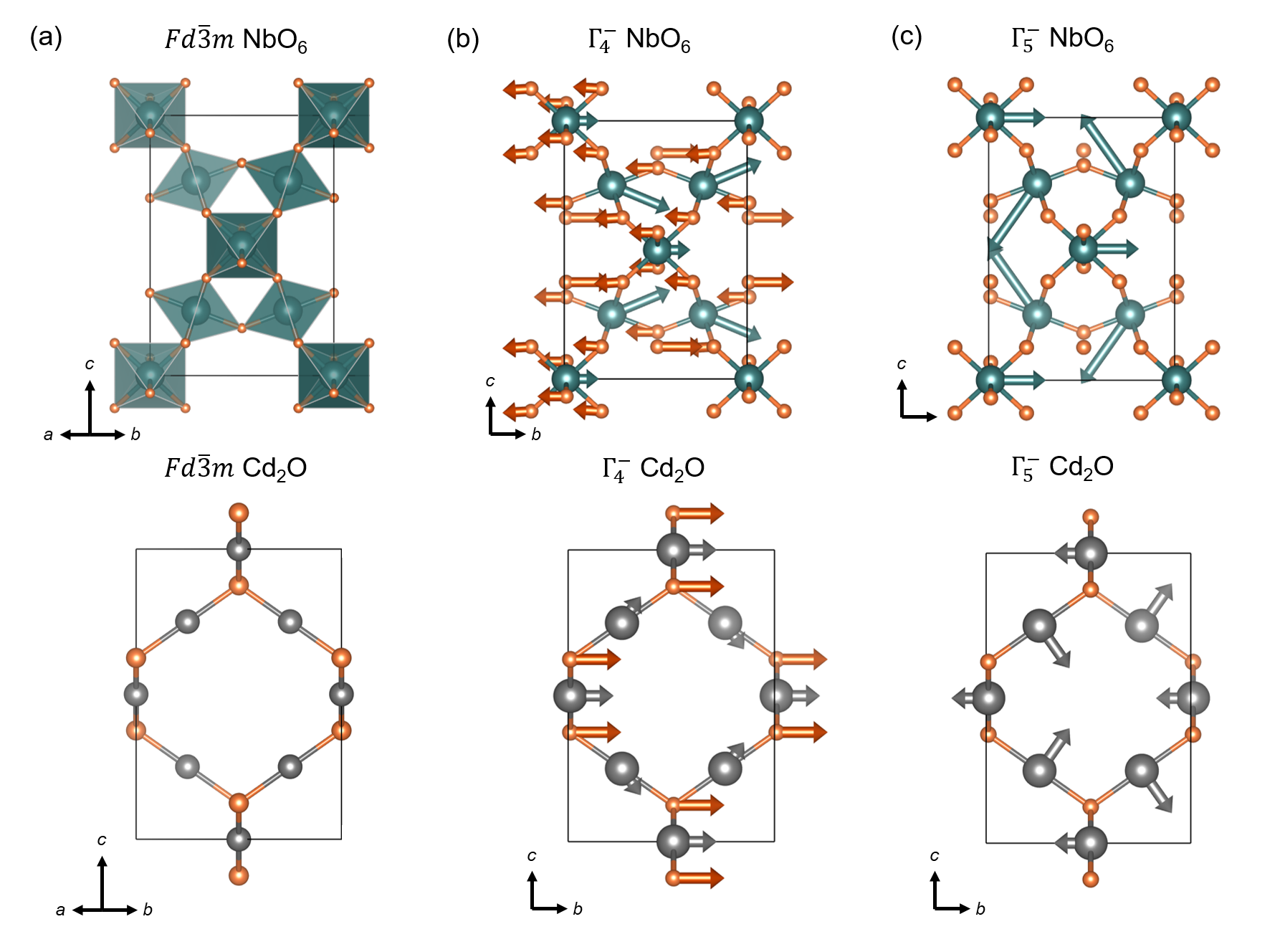}
	\caption{(a) NbO$_6$ and Cd$_2$O sublattice positions in the ideal $Fd\overline{3}m$ structure and subsequent (b) \GMIV and (c) \GMV mode distortions that lead to the low temperature $Ima2$-$\alpha$ polar structure. Adapted from Laurita \textit{et al.} \cite{laurita}.} 
	\label{Structure} 
\end{figure*}

A well-known example is the valence-precise pyrochlore \CNO, which undergoes a relaxor-type ferroelectric transition between 204\,K and 196\,K from a cubic $Fd\overline{3}m$ phase above 204\,K (\autoref{Structure}a) to a polar orthorhombic phase (point group $mm2$ \cite{ye}), well-described as having $Ima2$ symmetry \cite{malcherek,laurita}, below 196\,K. 
The intermediate phase between 204\,K and 196\,K has been challenging to identify because of its limited temperature range, but is believed to have an $mmm$ point group \cite{ye}, implying from group-subgroup relations that it has $Imma$ symmetry \cite{Tachibana2013}.  This is inconsistent with density functional theory (DFT) studies that find that the structural mode that defines $Imma$, $\Gamma_5^+$, is completely stable \cite{fischer,laurita} and so should not condense.

Comparison of the $Ima2$ structure with the cubic reference using group theory identifies \GMIV and \GMV as the dominant distortion modes describing the symmetry-lowering process (visualized in \autoref{Structure}b). Phonons calculated from first principles for the cubic phase likewise exhibit two soft phonon modes at $\Gamma$ with irreps matching the \GMIV and \GMV distortion modes present in the $Ima2$ structure. However, the exact nature of the ferroelectric phase transition, and the role of the two distortion modes, is unclear, and differing scenarios have been proposed \cite{buix}.

Furthermore, DFT identifies \GMIV as the more unstable mode, suggesting it is the primary driving mode, although the energy difference between the two is rather small \cite{fischer,laurita}. However, \GMV displacements are found to be dominant in global structural refinements, and exhibit a classic order-parameter-like behavior as a function of temperature, as opposed to the weak temperature dependence of \GMIV \cite{malcherek}. We therefore denote the \GMV-dominant phase revealed by the global refinement as $Ima2$-$\alpha$ and the \GMIV-dominant phase suggested by DFT as $Ima2$-$\beta$ in the remainder of the paper.

At temperatures below 100\,K, two further transitions occur to monoclinic phases, with likely space group $Cc$ \cite{Kuster1991,fischer}. Refinements to date have not been performed for these two monoclinic phases.  Again, DFT studies for the suggested $Cc$ space group indicate that \GMIV should be dominant.  Moreover, the phase transition picture is further complicated by the presence of several additional strain and displacive modes \cite{malcherek}, whose role in the stabilization of the various phases is poorly understood.

In this study, we find from an analysis of x-ray data a local structure that is dominated by \GMIV, but reduces to a dominant \GMV for the global structure. This finding is further supported by our DFT study.  Such `averaging out' of the polar \GMIV distortions over long distances is typical behavior for a relaxor ferroelectric.  Intriguingly, some of our local structural descriptions exhibit two-in/two-out behavior of the Nb displacements, as previously suggested by Malcherek based on simulations of diffuse scattering data \cite{malcherek2}.  Although those data suggest \CNO might be a long sought charge analog of pyrochlore spin ices \cite{Seshadri2006},  our results are more consistent with a Heisenberg-like rather than Ising description of the dipolar fluctuations.

\section{Methods}

\subsection{Density functional theory}

First-principles calculations were performed using DFT as implemented in the Vienna \textit{ab initio} simulation package (VASP) \cite{Kresse1996,Joubert1999} using a plane-wave basis set with a 750\,eV energy cutoff. The Perdew-Burke-Ernzerhof (PBE) exchange-correlation functional revised for solids (PBEsol) \cite{Perdew2007} was employed, along with the projector augmented wave (PAW) potentials to treat the separation of the core and valence electrons \cite{Blochl1994}. 

Energy surfaces were generated by perturbing the high-symmetry cubic experimental structure (as refined in Ref.~\cite{laurita}) along the soft eigendisplacements of the force constant matrix. These form two triply degenerate sets of orthogonal displacement vectors (exhibiting \GMIV and \GMV symmetry, respectively), each set of displacements forming a three-dimensional distortion space. These 3D distortion spaces were studied by perturbing the crystal structure along linear combinations of the basis eigenvectors and then relaxing the primitive cell size and shape until the forces and stress tensor were converged to 2 meV\,\AA$^{-1}$ and 5 meV\,\AA$^{-2}$, respectively.

\subsection{Sample preparation}

A polycrystalline sample of \CNO was prepared by mixing a stoichiometric ratio of CdO (Alfa Aesar, 99.998\%) and Nb$_2$O$_5$ (Alfa Aesar, 99.998\%). The sample was pressed into a pellet and fired in air at 500\,$^{\circ}$C for 10\, h to prevent Cd volatilization, followed by a subsequent grinding, pressing, and heating at 1000\,$^{\circ}$C for 12\,h.

\subsection{X-ray scattering}

X-ray pair distribution function (XPDF) measurements were performed at the National Synchrotron Light Source II, Brookhaven National Laboratory, on the Pair Distribution Function (PDF) beamline. The sample was ground into a fine powder and loaded into a 1 mm Kapton capillary for measurement. Data were collected upon cooling with a wavelength of 0.1665 \AA\/ every 20\,K from 300--80\,K. 2D scattering data were collected on a Perkin-Elmer flat panel amorphous Si-based area detector run at 4 Hz, with each dataset corresponding to the sum of 60 second data collection (240 frames).  Empty Kapton capillaries
were measured under identical conditions and used for background subtraction. GSAS-II \cite{Toby2013} was utilized to integrate the 2D data to 1D diffraction patterns. Corrections to obtain $S(Q)$ and subsequent Fourier transform with $Q_{max}=22$\,\AA$^{-1}$ and an $r$-grid of 0.01\,\AA\ were performed using the program PDFgetX3 \cite{Juhas2013} to obtain the x-ray pair distribution function $G(r)$. These parameters were chosen to optimize the $r$-resolution while minimizing Fourier termination ripples satisfactorily for all temperatures across the series. Least-squares refinement of the XPDF data were performed with PDFgui \cite{Farrow2007}. 

\section{Results and Discussion}

\subsection{Previous DFT results}

\begin{figure}
\centering
\includegraphics[width=1.0\columnwidth]{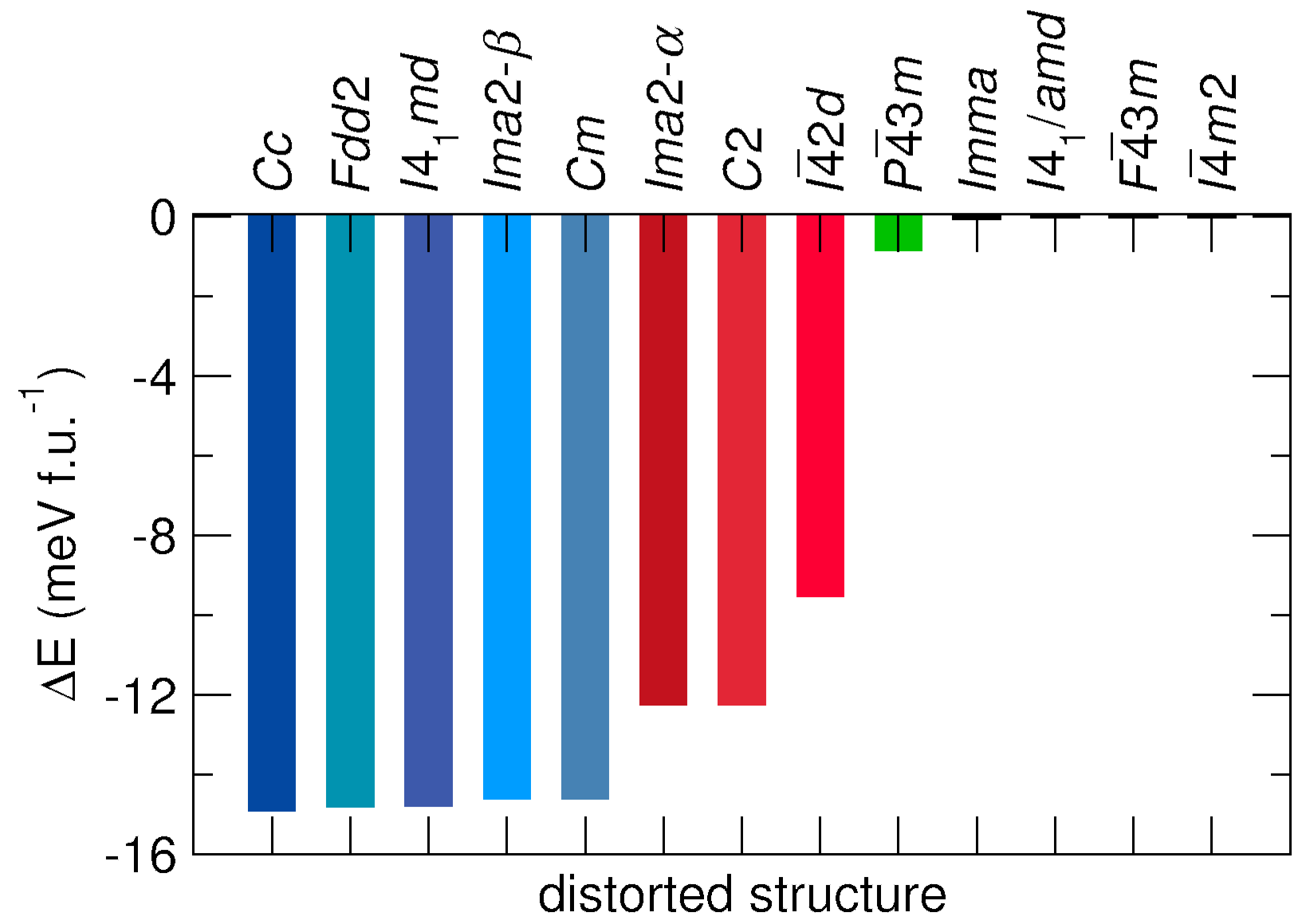}
	\caption{Calculated energies for \CNO in various space groups relative to the cubic phase. The bars colored in blue have a dominant \GMIV component, the ones in red a dominant \GMV component, and the variant in green is driven by an $X_4$ mode.  
	The four structures on the right without bars are associated with modes that are stable, and therefore do not lower the energy.  Updated from Laurita \textit{et al.} \cite{laurita}.
	}
	\label{Energy} 
\end{figure}

An important finding of previous DFT studies is that a number of space groups have almost identical energies (\autoref{Energy}), with the \GMIV-dominated ones having an energy only a few meV per formula unit below those of the \GMV-dominated ones, which are in turn also close in energy \cite{laurita}.  
This implies almost isotropic energy surfaces for both modes, which are also coupled to one another.  
The Landau theory for this has been written out in our previous study \cite{meier2022}, where we also illustrated the existence of various Higgs and Goldstone modes, and their anti-phase variants (Leggett modes), that should exist as a consequence.  The former are also found to play an important role in the related pyrochlore Cd$_2$Re$_2$O$_7$ \cite{kendziora,venderley}.

\begin{figure*}
\centering
\includegraphics[width=2.0\columnwidth]{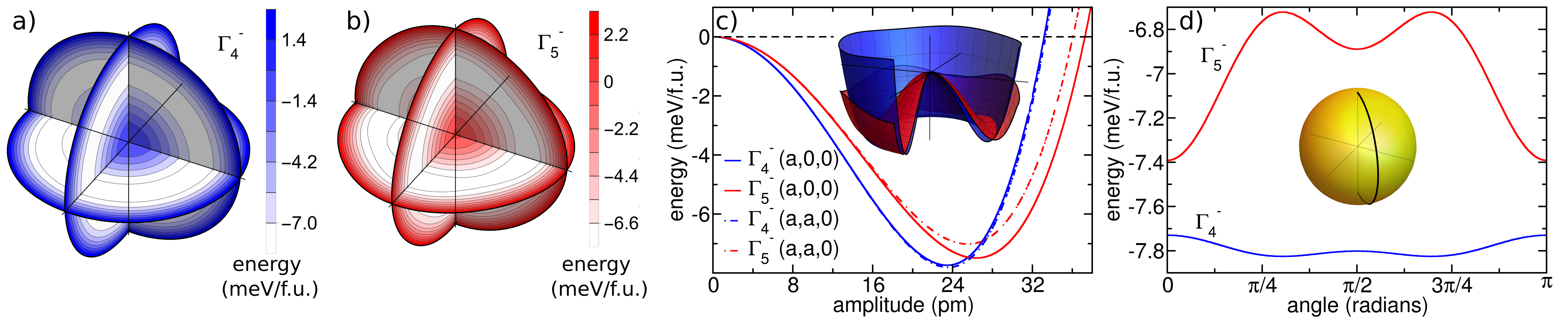}
	\caption{(a,b,) Contour maps of the energy surfaces for the three-dimensional (a) \GMIV and (b) \GMV distortion modes. (c) Energy reduction relative to the cubic phase as a function of distortion amplitude along the (a,0,0) (solid curves) and (a,a,0) (dot-dashed curves) order parameter directions for \GMIV and \GMV. The inset shows the two-dimensional energy surfaces defined by the order parameter (a,b,0) for \GMIV and \GMV. (d) Energy reduction relative to the cubic phase with constant distortion amplitude, r, varying the order parameter direction for \GMIV and \GMV through one half of a great circle, that is (a,a,b) with 2a$^2$+b$^2$=r$^2$, following the path shown by the black curve on the sphere in the inset.
	}
	\label{fig:energy_surface} 
\end{figure*}

\subsection{Local versus global structure}

This brings up the question of the nature of the phase transitions at 204\,K and 196\,K (it is below the latter temperature that relaxor behavior is most apparent).  It has been proposed that in the ferroelectric phase, there is a mixture of order-disorder and displacive behavior, with experimental results indicating that Nb is primarily displacive, whereas Cd is primarily order-disorder \cite{Pasciak2010,malcherek,malcherek2}. 
Alternatively, \GMV may be more displacive-like while \GMIV may be more order-disorder-like.  This highlights the fact that the DFT analysis is for a homogeneous material, whereas relaxor behavior implies local inhomogeneities.

To address these points, we note that while crystallographic analysis (such as Rietveld refinement) can provide a detailed description of global order and phase transitions as a function of temperature, local probes are necessary to reveal any local distortions or mid-range ordering that might average out in a global refinement. 
Techniques such as IR and Raman spectroscopy are valuable for describing the coordination environment of the cations, but are unable to describe any structural ordering between next-nearest neighbors, such as Cd-Cd/Nb-Nb ($M-M$) and Cd-Nb ($M-M^\prime$) interactions. 
The pair distribution function (PDF) is obtained from a combination of Bragg and diffuse scattering events when a material is irradiated, and is a histogram of all atom-atom interactions in a material, presented as a function of $r$-range. This technique is similarly useful for describing coordination environments (typically captured in peaks below 3\,\AA\/), and can additionally provide insight on mid- to long-range atomic correlations. 
Local ordering is often hidden as structural disorder in the form of enlarged atomic displacement parameters in a Rietveld refinement, but can have meaningful impact on the observed behavior of a material, such as relaxor behavior. In this study, `local' correlations were fit over a range of 1.7 - 5.7\,\AA\/, `mid-range' correlations were fit over a range of 1.7 - 12.0\,\AA\/, and `long-range' or `average' correlations were fit over a range of 5.7 - 30\,\AA\/. 
For our study, x-ray scattering was chosen as a probe due to the high neutron absorption cross-section of Cd (2520 barn) \cite{NIST_xs}. It should be noted that this necessary choice unfortunately de-emphasizes contributions from oxygen, which is a poor scatterer of x-rays in comparison to the heavy metals in \CNO; however, meaningful information about Cd-Cd/Nb-Nb ($M-M$) and Cd-Nb ($M-M^\prime$) interactions can be obtained, as well as some information on Cd-O and Nb-O bond interactions. 
Data were collected from room temperature to 80K to elucidate any local behavior in the cubic and orthorhombic phases, and as such our analysis will not experimentally address the lower temperature monoclinic phases mentioned above.

\subsection{Local \GMIV displacements: DFT simulations}

Before analyzing the coupling and competition between the two dominant distortion modes, we consider the energy landscape associated with each mode independently. 
Using the eigendisplacements of the force constant matrix as a basis set, we compute the energy surfaces for pure \GMIV and pure \GMV distortions in three dimensions (\autoref{fig:energy_surface}). Corresponding order-parameter directions and symmetries for the modes are given in \autoref{table1}. 
We find that both energy surfaces are remarkably isotropic, resulting in a 3D version of the well known `Mexican hat' potential that describes the corresponding 2D case. 
Pure displacements associated with the order parameter direction along the cubic axes are slightly more energetically favorable than off-axis directions for \GMV, but for the most part a displacement of a given magnitude lowers the energy to the same degree regardless of the order parameter direction (\autoref{fig:energy_surface}c). 
The particularly high degree of \GMIV isotropy is consistent with polarization measurements near the ferroelectric transition temperature, which found that the polarization direction is easily rotated by an electric field \cite{ye}. Note that these calculations describe the 0\,K state, and we might expect thermal energy to further reduce the weak anisotropy. Indeed, the barrier between energy minima is less than 0.5 meV/f.u.\ for \GMV and less than 0.1 meV/f.u.\ for \GMIV (\autoref{fig:energy_surface}d), implying fluctuations between order parameter directions should readily take place. A similar behavior for the energy surface of \GMIV has been remarked on for perovskite relaxors \cite{Heitmann2014}.  

\begin{table}
\caption{Order parameter directions for space groups within the manifold of $\Gamma_4^-$, $\Gamma_5^-$ modes (generated using ~\onlinecite{byu}).}
\begin{ruledtabular}
\begin{tabular}{lll}
$\Gamma_4^-$  & $\Gamma_5^-$ & Space Group \\
\hline
(a,0,0) &  & $I4_1md$ \\
  & (a,0,0) & $I\overline{4}2d$ \\
(a,0,0) & (0,b,0) & $Fdd2$ \\
(a,a,0) & (0,b,-b) & $Ima2$ \\
(a,b,0) & (0,c,d) & $Cc$ \\
(a,a,b) & (0,c,-c) & $Cm$ \\
\end{tabular}
\end{ruledtabular}
\label{table1}
\end{table}

\begin{figure*}
\centering
\includegraphics[width=2.0\columnwidth]{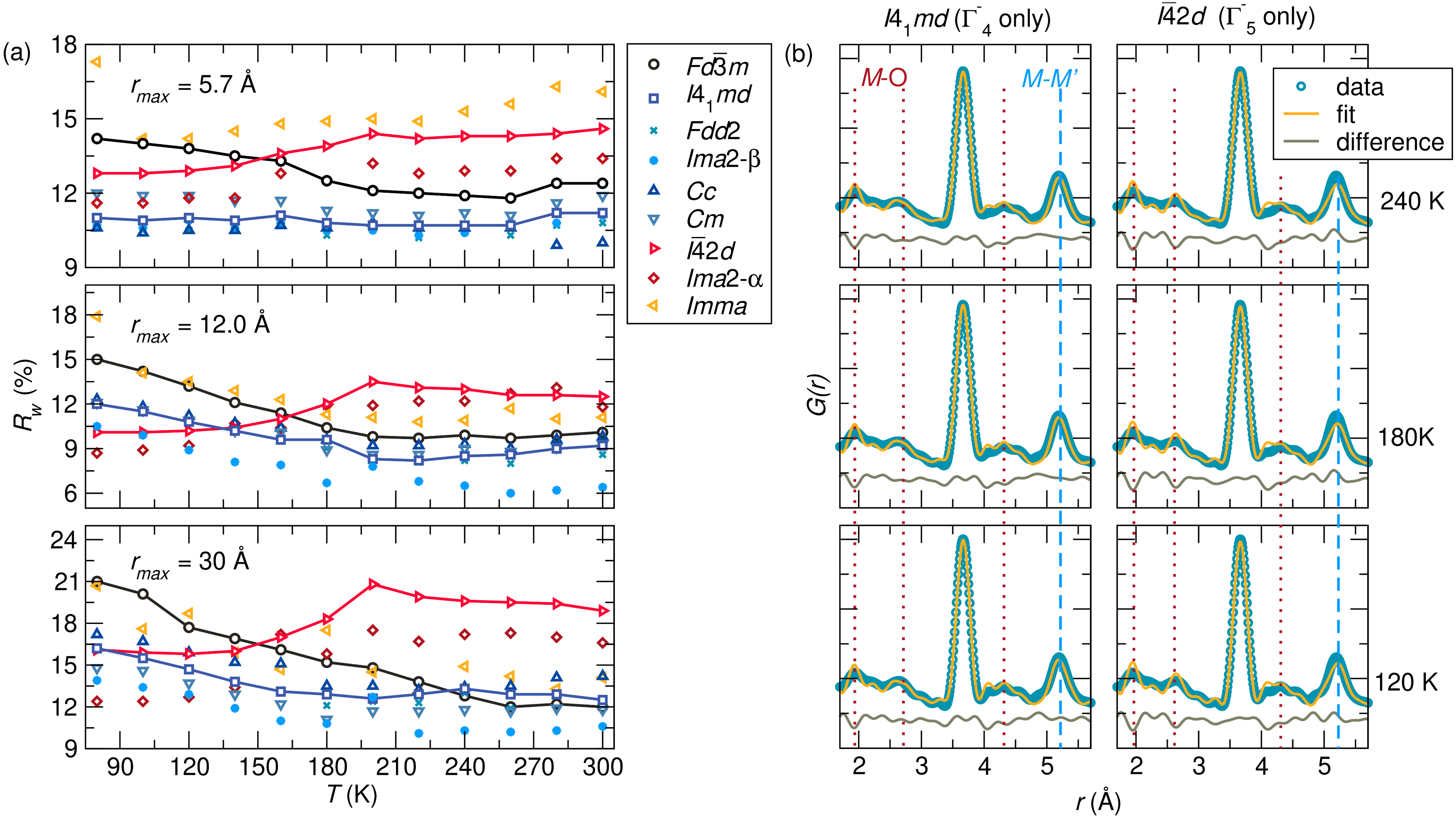}
	\caption{(a) Goodness of fit ($R_w$) as a function of temperature for a series of candidate space groups with a fit $r_{max}$ of 5.7\,\AA\/,  12.0\,\AA\/, and 30\,\AA\/. The fit against the cubic $Fd\overline{3}m$ structure is shown as black circles, space groups with primarily \GMIV distortion modes are shown as blue symbols, space groups with primarily \GMV distortion modes are shown as red symbols, and the space group with a primary $\Gamma$$_5$$^+$ distortion mode ($Imma$) is shown as yellow symbols. (b) Local fits (1.7--5.7\,\AA\/) of the XPDF data against the $I4_1md$ and $I\overline{4}2d$ space groups at 240\,K, 180\,K, and 120\,K. The $M$-O correlations are indicated by the red short-dash lines, and the $M$-$M'$ correlations by the blue long-dash lines.}
	\label{PDF_combined} 
\end{figure*}

 Whereas the energy scales are comparable, we nonetheless observe that the \GMIV mode is both more unstable (\textit{i.e.\@\xspace}, it has a more negative curvature of the energy versus displacement around zero amplitude) and has a lower energy minimum than \GMV, consistent with previous first principles investigations  \cite{fischer,laurita}. 
 Until now, this has been perceived as inconsistent with experimental measurements, where \GMV displacements dominate the global refinement \cite{malcherek}. 
However, the isotropic energy surfaces of the two modes imply that long-range characterization techniques may be blind to the local mode dynamics along the energy surfaces \cite{meier2022}. We also observe that the \GMIV energy surface is significantly more isotropic than the \GMV one (\autoref{fig:energy_surface}c,d), implying that the \GMIV distortions exhibit a greater degree of disorder at higher temperatures.  Conversely, the larger anisotropy of the \GMV energy surface enables \GMV to condense over longer length scales than \GMIV, consistent with the dominance of the \GMV mode in global refinements of the crystallographic structure.

\subsection{Local \GMIV displacements: X-ray scattering}

To investigate this further, we performed XPDF measurements as a function of temperature and fit the results with a variety of space groups suggested from the calculated energy surface as possible descriptions of the local symmetry breaking.
Eight candidate models with decreasing symmetry/ordering were chosen to fit the data, falling into four categories: 
(1) the high temperature cubic $Fd\overline{3}m$ structure, (2) five candidate structures dominated by the \GMIV distortion mode, (3) two candidate structures dominated by the \GMV distortion mode, and (4) an orthorhombic $Imma$ structure previously suggested as the intermediate phase between $Fd\overline{3}m$ and $Ima2$ (for (2) and (3), see \autoref{table1}). 
All structures were derived from our previous work on \CNO \cite{laurita}, except for the \GMIV-dominated $Cc$ structure which was taken from Ref.~\cite{fischer} and further relaxed with DFT. To minimize the number of refined parameters in the least-squares analysis of the PDF, atomic positions were fixed to the theoretical atomic positions obtained from the DFT analysis. 
Due to the minimal x-ray scattering power of oxygen in the presence of heavier cations, the oxygen atomic displacement parameters (ADPs) were modeled as being isotropic (U$_{iso}$) with a fixed value of 0.01\,\AA\/$^2$. For these fits, the scale, lattice parameters, and isotropic Cd and Nb ADPs were allowed to refine, and a correlated motion correction ($\delta$-2) was applied to the fits to account for peak sharpening at low-$r$. Goodness of fit (reported as $R_w$) as a function of temperature for each candidate space group is shown in \autoref{PDF_combined}.

\begin{figure*}
	\centering \includegraphics[width=1.5\columnwidth]{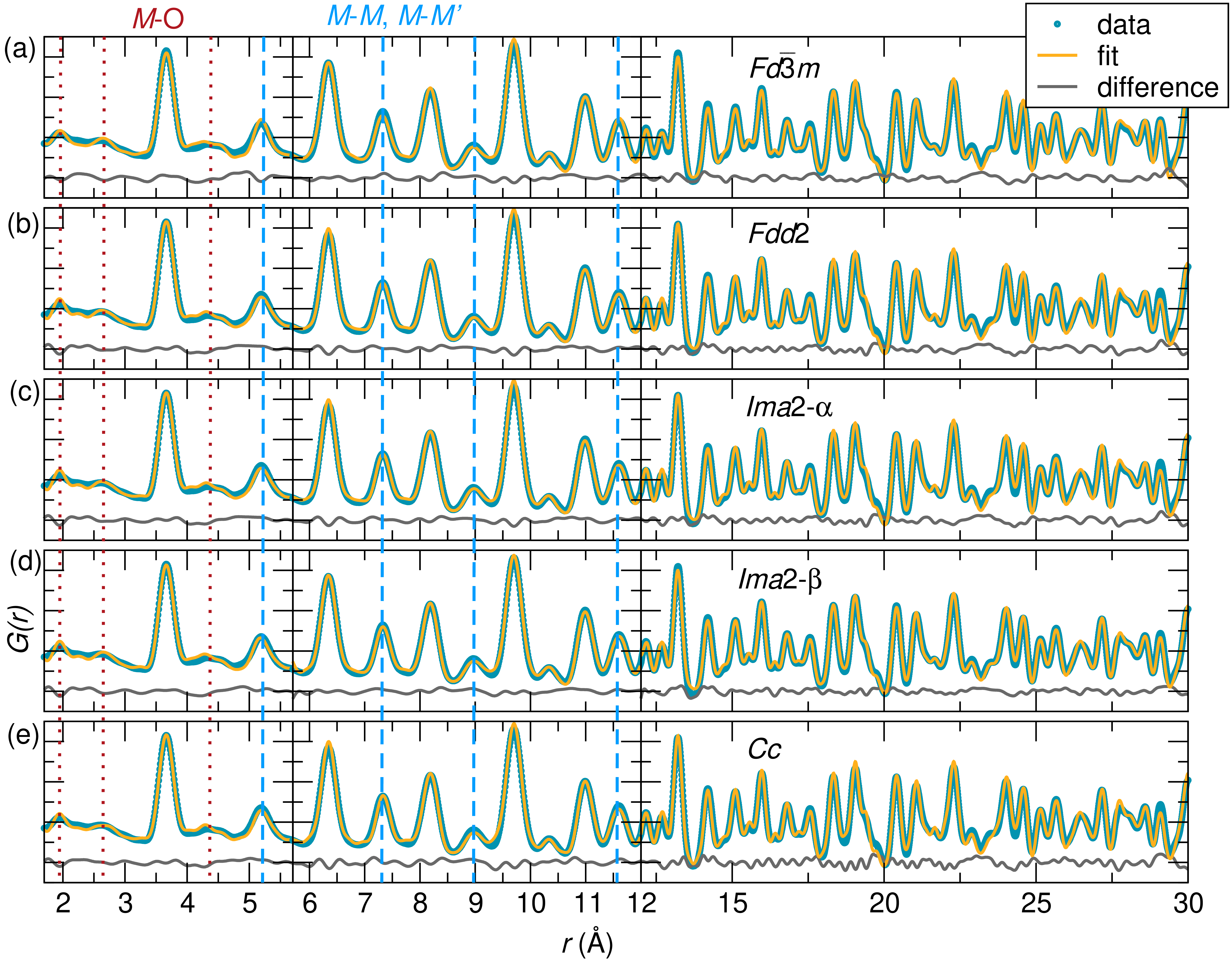}
	\caption{Fits of the XPDF data collected at 300\,K against proposed structures for Cd$_2$Nb$_2$O$_7$: (a) cubic $Fd\overline{3}m$, (b) orthorhombic $Fdd2$, (c) orthorhombic $Ima2$-$\alpha$, (d) orthorhombic $Ima2$-$\beta$, and (e) monoclinic $Cc$. Fit ranges are indicated by the range shown in each panel, $M$-O correlations of interest are shown as short red dashes, and $M$-$M$/$M$-$M^\prime$ correlations of interest are shown as long blue dashes.} 
	\label{Fig1_300K_fits} 
\end{figure*}

From a long-range perspective (correlations up to 30 \AA), the space groups dominated by \GMIV distortions better fit the data at high temperature, with a cross-over to structures dominated by the \GMV distortion mode around 140\,K (bottom graph of \autoref{PDF_combined}a). 
This length scale provides a description of the data closer to the crystallographic structure (\textit{i.e.\@\xspace} encompassing several degrees of atom-atom interactions), but does not capture any microstructural effects that might be at play, such as the role of nanodomains in the material. Such analysis would be better suited to microscopy studies \cite{Stemmer}, which are outside of the scope of the present work.

It should be noted that, whereas the XPDF fits provide insight into the prevalence of \GMIV versus \GMV -type local distortion patterns, it is important to be conservative in stating which specific structure is the best description.  For instance, the $Cm$ space group has nine refined parameters as opposed to three in $Fd\overline{3}m$, five in $Fdd2$, seven in $Ima$2, and eight in $Cc$. Nonetheless, since all \GMIV-based structures have a lower $R_w$ than both of the \GMV\ space groups, we can say that the \GMIV distortion pattern better describes the data over this fit range.

At 300\,K, both cubic and \GMIV-type models equally describe the data out to 30\,\AA\/, with the highest symmetry cubic $Fd\overline{3}m$  having larger atomic displacement parameters (ADPs) and the lowest symmetry $Cm$ structure having small distortions in the atomic positions, suggesting that the cubic and \GMIV-type models are equivalent descriptions at this length scale. However, both the enlarged anisotropic ADPs in the ideal $Fd\overline{3}m$ model and atom off-centering in the \GMIV models indicate that distortions are present and cations are not located at the center of their coordination environments. In the mid-range region (correlations up to 12 \AA, middle graph of \autoref{PDF_combined}a), the trend is similar to the long-range fits, with $Ima2$-$\beta$ providing a better description of the mid-range structure upon cooling to 120\,K, whereas $Ima2$-$\alpha$ becomes the better description of correlations at this length scale for lower temperatures.  In the local range (correlations up to 5.7 \AA, top graph of \autoref{PDF_combined}a), the space groups with primarily \GMIV distortion modes provide a better fit of the data at all temperatures. $Cc$ has the lowest $R_w$ at 300 K and exhibits similar $R_w$ values to $Cm$, $Ima2$-$\beta$, and $Fdd2$ upon cooling. 

To better understand the nature of the distortions, comparisons of the fits to the $I4_1md$ (\GMIV only) and $I\overline{4}2d$ (\GMV only) structures above (240\,K), near (180\,K), and below (120\,K) the ferroelectric transition were performed. The local fits (1.7--5.7\,\AA\/) to the XPDF are shown in Figure \ref{PDF_combined}b. At all temperatures, $I4_1md$ provides a better fit than $I\overline{4}2d$ of the Cd-O peaks (Cd coordination environment) and the $M-M^\prime$  peak at approximately 5.2\,\AA\/, which is the next-nearest Cd-Nb correlation.
Moreover, at all temperatures, $Fdd2$ further improves the fit of all $M$-O peaks. $Ima2$-$\beta$ has a similar improvement in the $M$-O local peaks, in addition to a better fit of the mid-range $M-M$ and $M-M^\prime$ peaks. This is illustrated for 300 K data in \autoref{Fig1_300K_fits}.

\subsection{Intermediate phase scenario}

Neither our first-principles results nor our experimental data provide strong evidence of an intermediate phase between 204\,K and 196\,K belonging to point group $mmm$ before the relaxor ferroelectric transition occurs. We calculate no unstable or low-lying phonon modes in the high-symmetry structure which break the requisite symmetries, and XPDF data do not favor the $Imma$ structure at any length scale.
Based on the above results, we can offer a possible scenario for the intermediate phase.  Our DFT results indicate that $Cm$, $Cc$, $Ima2$-$\beta$, $Fdd2$, and $I4_1md$ structures have almost identical energies (\autoref{Energy}). The last two only differ in that $I4_1md$ has no \GMV component.  To understand the proposed scenario, note that $Fdd2$, $Ima2$ and $Cc$ define a great circle on each of the \GMIV and \GMV energy surfaces (\GMIV and \GMV are each three-dimensional irreps, see \autoref{fig:energy_surface}), with $Fdd2$ corresponding to $\phi=0^\circ$, $Ima2$ to $\phi=45^\circ$, and $Cc$ to an intermediate angle, as can be seen from \autoref{table1} \cite{meier2022}.
This is exactly analogous to Cd$_2$Re$_2$O$_7$, where the respective space groups are $Ima2$, $I4_122$ and $F222$.  For the latter, one has a proposed sequence of transitions from $Fd\bar{3}m \rightarrow I4m2 \rightarrow I4_122 \rightarrow F222$ \cite{Kapcia2020}, which  corresponds to various positions around the brim of the `Mexican hat' energy surface (in this case, from $\Gamma_3^-$ which is a two-dimensional irrep).
The analogous series for \CNO would be $Fd\bar{3}m \rightarrow Fdd2 \rightarrow Ima2 \rightarrow Cc$.
There are several points in favor of this scenario, in that it explains the experimental absence of the (0,0,10) Bragg peak in the temperature range of 196K to 204K \cite{Tachibana2013}, since this is allowed for $Ima2$ but not for $Fdd2$.  It also explains why optical studies indicate a ferroelectric polarization vector along the cubic axis \cite{ye} as opposed to $Ima2$ where it is along (1,1,0) instead.  $Fdd2$ is also found to be the global space group upon sulfur doping \cite{laurita} and has been proposed to be the local structure for \CNO itself \cite{malcherek2}.  The claim of a centrosymmetric space group ($mmm$ point group) in this temperature range \cite{ye} could then be a domain averaging effect, similar to the phenomenon observed in cubic BaTiO$_3$ \cite{Tsuda2016}.

\subsection{Cd and Nb displacements}

Our results help to explain the challenges associated with identifying possible intermediate space groups. The isotropic energy surfaces (\autoref{fig:energy_surface}) imply that local distortions along multiple order parameter directions may occur simultaneously, a result corroborated by the number of space groups with comparable goodness of fit in the XPDF data. The observation that local fits differ significantly from the mid- and long-range fits also supports the conclusion that local structural disorder (in the form of uncorrelated atomic displacements) plays a large role in the relaxor ferroelectric transition of \CNO.

The combination of DFT and our XPDF data implies a distinction between the role of the two primary distortion modes. \GMIV displacements dominate the local structure even at high temperatures, whereas \GMV displacements condense over longer length scales at lower temperatures, consistent with previous global refinements \cite{malcherek}. 

\begin{figure}
\centering
\includegraphics[width=1.0\columnwidth]{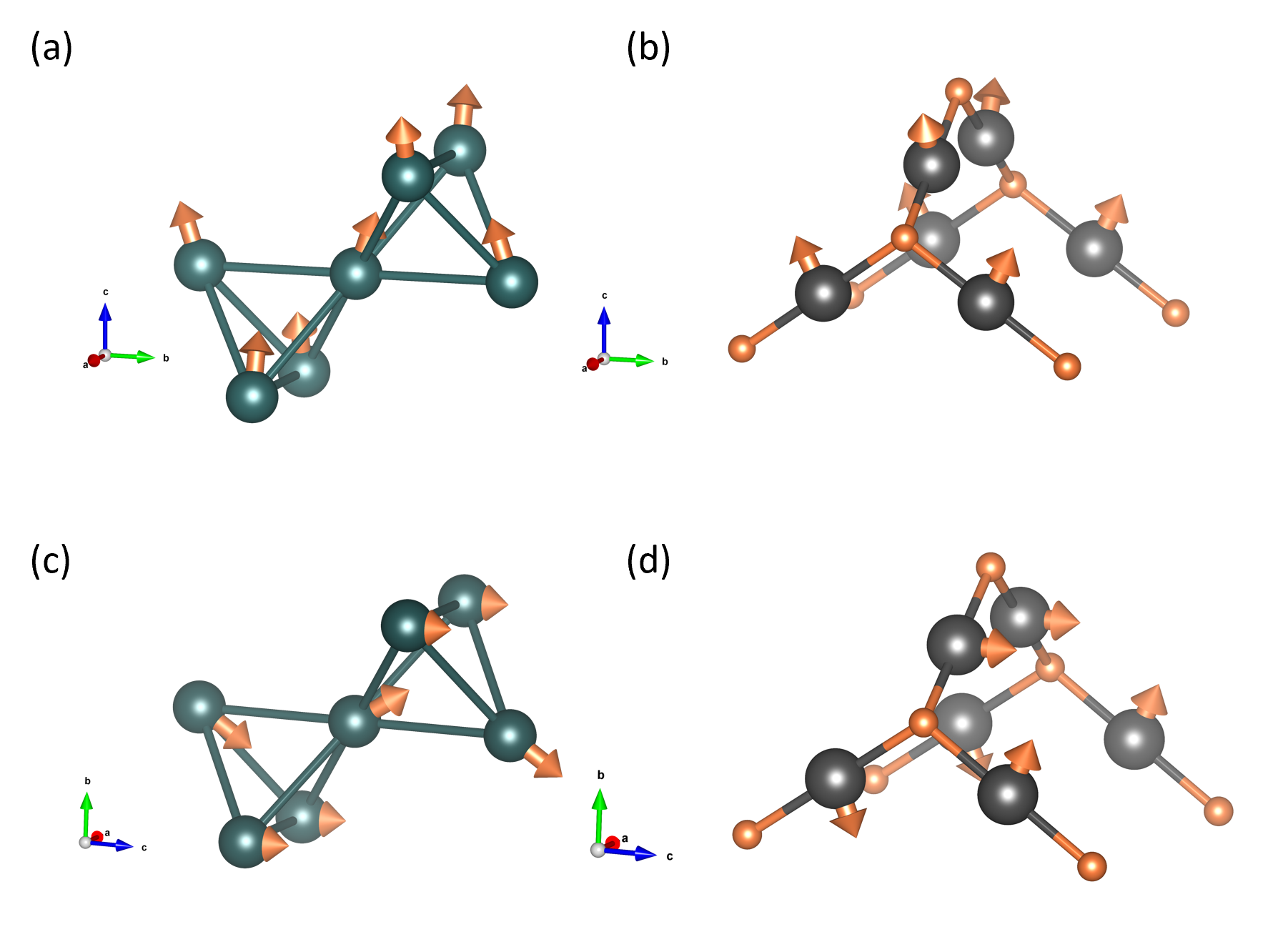}
	\caption{ (a,b) Displacement pattern of the relaxed $Fdd2$ structure relative to the high-symmetry cubic structure ($Fd\overline{3}m$) for the (a) Nb sublattice and (b) Cd-O sublattice. (c,d) Displacement pattern of the relaxed \GMIV-dominated $Ima2$-$\beta$ structure relative to the high-symmetry cubic structure ($Fd\overline{3}m$) for the (c) Nb sublattice and (d) Cd-O sublattice. }
	\label{fig:displacements} 
\end{figure}

Visualizing the \GMIV and \GMV displacement patterns (\autoref{fig:displacements}) also allows us to comment on the relative role of Cd and Nb in the phase transitions. As Nb displacements (likely due to the second-order Jahn-Teller effect typical of $d^0$ ions) are central to the ferroelectric behavior, we turn to a detailed description of these.  
As commented on by Malcherek \cite{malcherek2}, $Fdd2$ is a possible description of the local structure. This is not inconsistent with our analysis, in that $Fdd2$ produces one of the lowest $R_w$ values at the local scale below 270K (\autoref{PDF_combined}a) and corresponds to a local energy minimum for the \GMIV mode (see \autoref{table1} and \autoref{fig:energy_surface}). The $Fdd2$ displacements correspond to the classic two-in/two-out behavior characteristic of spin moments in pyrochlore spin ices.  That is, for a given tetrahedron of Nb, two of the Nb displacements point inwards, two point outwards (\autoref{fig:displacements}), with the Nb displacements along the Nb-Nb bond directions as opposed to along the local trigonal axis as in spin ices (here, the polar displacement along $c$ drops out, since all Nb ions are shifted by this).  A similar picture has been advocated for Bi displacements in Bi$_2$Ti$_2$O$_7$ \cite{Seshadri2006}. This implies that some of the same physics found in spin ices might be relevant for the dielectric response for \CNO.  In particular, violation of the ice rules leads to the creation of monopole-antimonopole pairs (\textit{i.e.\@\xspace}, three-in/one-out--three-out/one-in), whose dynamics can explain the Vogel-Fulcher relation for the relaxation rate.  
We note that monopole-antimonopole pairs have also been invoked in the context of cubic perovskite relaxors, where they represent eight-in--eight-out instead \cite{Nahas2016}.

However, as noted above, assignment of a specific local space group on the basis of our XPDF data is inadvisable. For example, XPDF fits against $Ima2$-$\beta$ (\GMIV-dominated) can produce $R_w$ values even lower than $Fdd2$. This can be attributed in part to the larger number of structural parameters in $Ima2$-$\beta$ than $Fdd2$, but nevertheless we cannot justify ascribing the local structure to one or the other. Similarly, the energy difference between $Fdd2$ and $Ima2$-$\beta$ as computed with DFT is below 1 meV/f.u., meaning neither can be called more energetically favorable. Therefore, while a charge ice description of \CNO is attractive, our results indicate a more Heisenberg-like description of the dipolar fluctuations than the Ising-like one implied by a charge ice picture. That is, our data depict an isotropic, \GMIV-dominated energy surface (\autoref{fig:energy_surface}), implying that Nb atoms displace within their local coordination environments along various order parameter directions within the \GMIV distortion space.

We now turn to Cd displacements. Cd has two short bonds with oxygen along its local trigonal axis (O(2) ions) and six long bonds with oxygen in the NbO$_6$ octahedra (O(1) ions).  To improve bonding with the O(1) ions, Cd displaces normal to the local trigonal axis (\autoref{fig:displacements}), as is seen in many pyrochlores. For instance, diffuse scattering in La$_2$Zr$_2$O$_7$ finds a pronounced annular displacement of La normal to the local trigonal axis \cite{Tabira2001}, consistent with what we find for Cd in \CNO, and Cd likely remains disordered in the ferroelectric phase \cite{Pasciak2010}. As such, Cd displacements are unlikely to drive the symmetry-lowering either locally or over longer ranges.

\section{Conclusion}

We have presented evidence that \CNO is locally characterized by the ferroelectric mode \GMIV but globally described by the non-polar mode \GMV.  This observation relates this pyrochlore to perovskite relaxor ferroelectrics whose relaxor behavior is attributed to local polar clusters \cite{Cowley2011}.  Recently, further insights into relaxor behavior in perovskites have been achieved by contrasting XPDF, which emphasizes higher $Z$ ions, with neutron PDF where oxygens play a significant role \cite{Krogstad}.  A neutron PDF analysis in \CNO would be possible with Cd isotopically-enriched samples. This could be complemented by a DFT determination of the various parameters in the Landau free energy that has been critical in studies of structural phase transitions, such as in the perovskite YMnO$_3$ where the amplitude of the structural order parameter is order-disorder-like but the phase behaves displacively \cite{artyukhin,Skjaervo2019,meier}, a study we hope to report on in the future.

In addition, we have presented candidate descriptions of the local structure, some of which ($Fdd2$ and $I4_1md$) are dominated by two in/two out displacements of Nb ions in a given Nb tetrahedron, which might imply that \CNO is a charge ice analog of the pyrochlore spin ices that have been much studied in the past two decades.  Our results, however, are consistent with a richer energy landscape that is more Heisenberg-like than Ising-like.  Therefore, inspired by recent work in spin ices \cite{Kassner2015,Eyvazov2018,samarakoon2021structural}, we believe that the dynamics of the dielectric response of \CNO is worth re-investigating.

\begin{acknowledgments}
D.H.Y and J.M.R. were supported by the National Science Foundation (NSF) under award number DMR-2011208.
G.L. acknowledges support for this work from Bates College, and from the National Science Foundation (NSF) through DMR 1904980.
M.R.N. was supported by the Materials Sciences and Engineering Division, Basic Energy Sciences, Office of Science, U.S. Department of Energy.
 N.A.S. was supported by the European Research Council (ERC) under the European Union’s Horizon 2020 research and innovation program project HERO grant (No. 810451) and by ETH Zurich.  Q.N.M. was supported by the Swiss National
Science Foundation under Project No. P2EZP2\_191872.
 This research used the Pair Distribution Function (PDF) beamline of the National Synchrotron Light Source II, a U.S. Department of Energy (DOE) Office of Science User Facility operated for the DOE Office of Science by Brookhaven National Laboratory under Contract No. DE-SC0012704. Calculations were performed using the Department of Defense High
Performance Computing Modernization Program (DOD-HPCMP).
 
\end{acknowledgments}

\bibliography{references.bib}

\end{document}